\documentclass{article}

\usepackage[utf8]{inputenc} 
\usepackage[T1]{fontenc}    
\usepackage[colorlinks,citecolor=blue!70!black,linkcolor=blue!70!black,
urlcolor=blue!70!black]{hyperref}       
\usepackage{url}            
\usepackage{booktabs}       
\usepackage{amsfonts}       
\usepackage{nicefrac}       
\usepackage{microtype}      
\usepackage{xcolor}         
\usepackage{wrapfig}
\usepackage[rightcaption]{sidecap}

\usepackage{fullpage}
\usepackage[ruled,vlined]{algorithm2e}
\usepackage{algorithm, algorithmic}

\usepackage{graphicx}
\usepackage{subfigure}
\usepackage{booktabs} 
\usepackage{mkolar_definitions}

\usepackage{enumitem}

\usepackage{multirow}

\allowdisplaybreaks
\usepackage{natbib}

\usepackage{xspace}
\usepackage{amsthm,amsmath,amssymb}
\usepackage{amsfonts,dsfont}
\usepackage{mathtools}
\usepackage{tikz}
\usepackage{comment} 

\usepackage{caption}

\newcount\Comments  
\definecolor{darkgreen}{rgb}{0,0.5,0}
\definecolor{darkred}{rgb}{0.7,0,0}
\definecolor{teal}{rgb}{0.3,0.8,0.8}

\newcommand{\kibitz}[2]{\ifnum\Comments=1\textcolor{#1}{\textsf{\footnotesize #2}}\fi}

\newcounter{relctr} 
\everydisplay\expandafter{\the\everydisplay\setcounter{relctr}{0}} 

\newcommand{\systemName}{EyeO}
\AtBeginDocument{} 

\title{EyeO: Autocalibrating Gaze Output with Gaze Input}
\author{
  Akanksha Saran\textsuperscript{1} \qquad Jacob Alber\textsuperscript{1}   \qquad Cyril Zhang\textsuperscript{1} \\ Ann Paradiso\textsuperscript{3} \qquad Danielle Bragg\textsuperscript{2}  \qquad John Langford\textsuperscript{1} \\
   \vspace{2mm} \\
  Microsoft Research NYC\textsuperscript{1} \qquad Microsoft Research New England\textsuperscript{2} \\ 
  Microsoft Research Redmond\textsuperscript{3}\\
\texttt{\{akanksha.saran, jacob.alber, cyrilzhang}\\
\texttt{annpar, danielle.bragg, jcl\}@microsoft.com} 
}

\usepackage{xspace}

\begin{document}

\maketitle
\begin{abstract}
Gaze tracking devices have the potential to greatly expand interactivity, yet miscalibration remains a significant barrier to use. As devices miscalibrate, people tend to compensate by intentionally offsetting their gaze, which makes detecting miscalibration from eye signals difficult. To help address this problem, we propose a novel approach to seamless calibration based on the insight that the system's model of eye gaze can be updated during reading (user does not compensate) to improve calibration for typing (user might compensate). To explore this approach, we built an auto-calibrating gaze typing prototype called \systemName{}, ran a user study with 20 participants, and conducted a semi-structured interview with 6 ALS community stakeholders. Our user study results suggest that seamless autocalibration can significantly improve typing efficiency and user experience. Findings from the semi-structured interview validate the need for autocalibration, and shed light on the prototype's potential usefulness, desired algorithmic and design improvements for users.
\end{abstract}

\section{Introduction}

Eye tracking technology has emerged as a valuable tool in a variety of applications, including accessibility, augmented reality (AR), virtual reality (VR), and gaming \citep{morimoto2005eye}. It has been particularly beneficial for individuals with motor impairments who rely on gaze-based input methods for communication and device control \citep{zhang2017smartphone, mott2017cascading}, including people with amyotrophic lateral sclerosis (ALS). In particular, gaze typing, a common use case of eye tracking, enables users to input text by looking at keys of an on-screen keyboard, thereby offering a hands-free and non-vocal method of communication.  
Beyond current applications, eye tracking technology has the potential to unlock additional interactions with the environments around us, for example by helping to better understand user attention and personalize displays accordingly.

Despite this potential, calibration difficulties remain a major barrier to use \citep{abdrabou2019calibration, kasprowski2018comparison, wang2016deep, bhatti2021eyelogin, kasprowski2016implicit, hiroe2023implicit}. Calibration is a process that establishes a relationship between the user's gaze and the corresponding screen coordinates (Fig. \ref{fig:calibration}). During calibration, the system typically guides the user through looking at a set of visual targets which enables an update to the internal model of the user's eye gaze. This process is time consuming and can be tedious and uncomfortable \citep{zhang2017smartphone}. 
Even after calibration, head movements, eye fatigue, changes in lighting and other environmental conditions, and hardware inconsistencies lead to miscalibration over time, which can significantly impact system performance \citep{jacob2003eye} and necessitate re-calibrations. This problem is further exacerbated for users with motor impairments who may have difficulty maintaining a stable head position or participating in traditional calibration procedures. 
Consequently, there is a need for calibration methods that are not only accurate but also adaptive to changes in the user and environment.

\begin{figure*}[ht]
  \centering
  \subfigure[Gaze providing system input: the user compensates for miscalibration to the right, intentionally focusing to the left of the key they want to select.]{
  \includegraphics[width=0.4\linewidth]{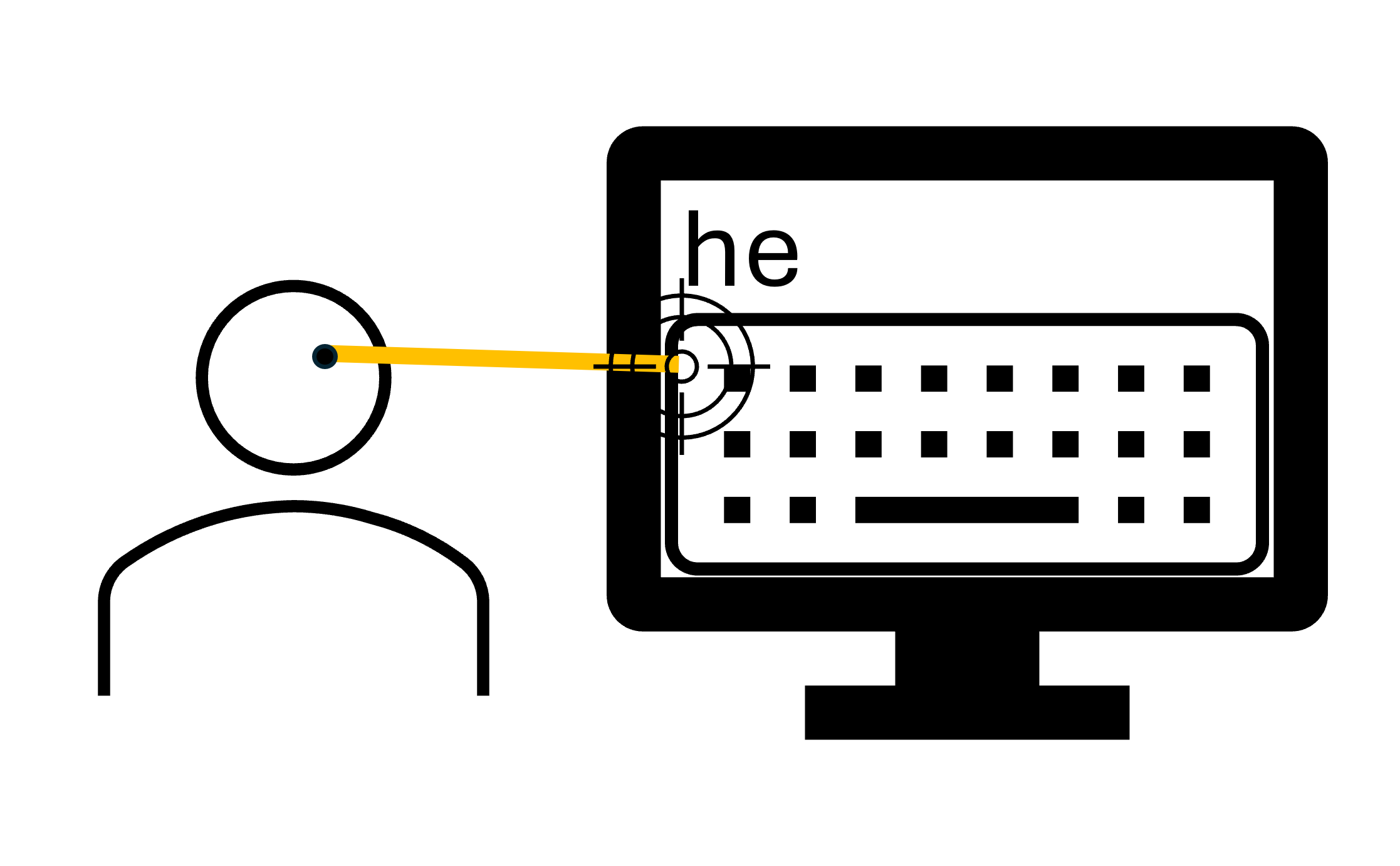}
  }
  \hspace{2mm}
  \subfigure[Gaze consuming system output: the user does not compensate for miscalibration, looking directly at the text they read. Their gaze does not control the system, so they are free to look where is natural.]{
  \includegraphics[width=0.4\linewidth]{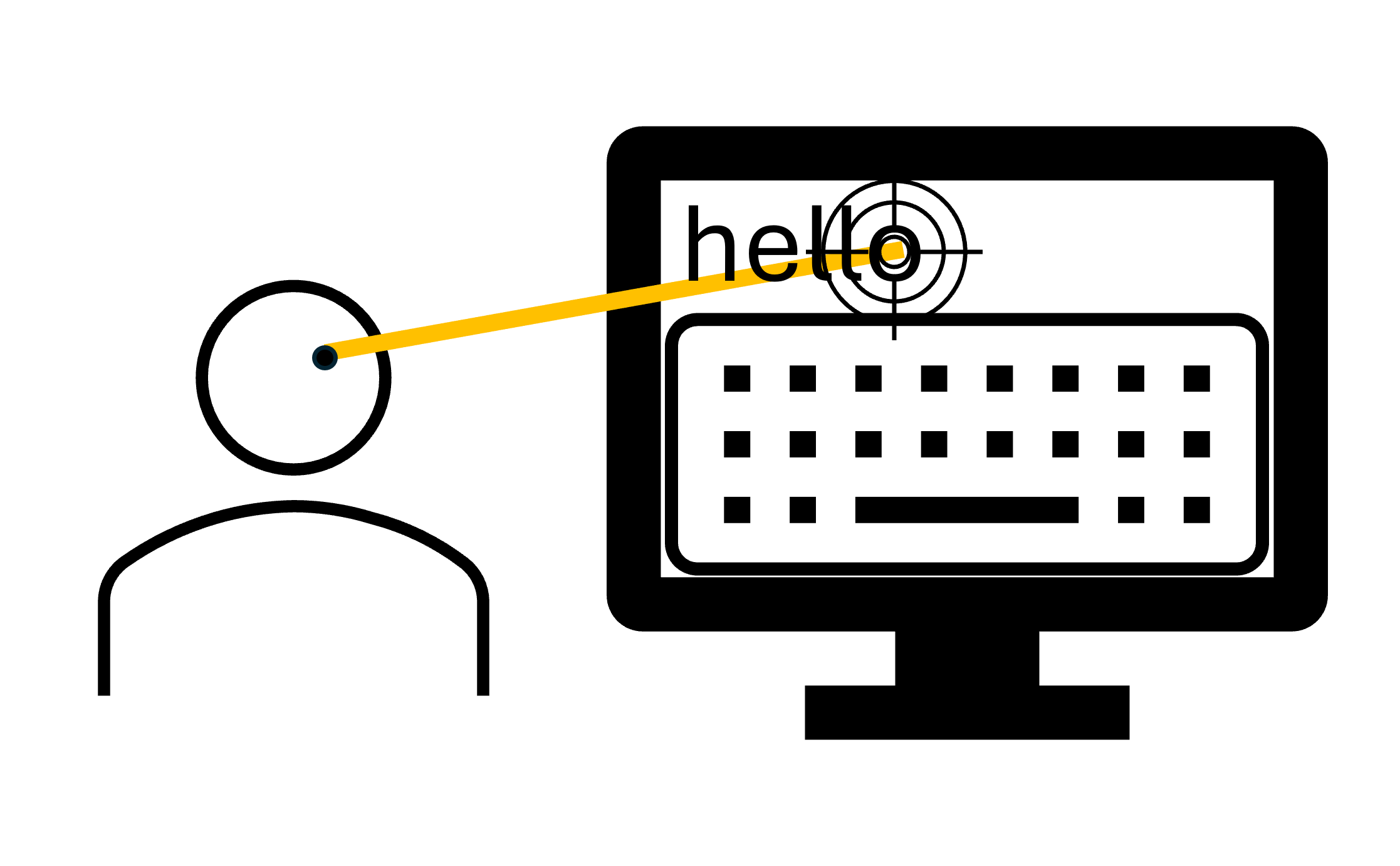}
  }
  \caption{Demonstration of the difference in gaze offset between system input (i.e. typing) and consumption of system output (i.e. reading). We leverage this difference to detect and correct miscalibrations, providing the user with a more seamless experience.}
  \label{fig:insight}
\end{figure*}

Automatically detecting and correcting for miscalibration could address the calibration problem, but is confounded by users compensating for miscalibrations. 
For example, when typing on dwell-based visual keyboards, users may compensate and autocorrect for miscalibration in eye tracking software by purposefully glancing at adjacent keys to activate keys of interest. While such behavior adds additional cognitive load for users, it can enable them to type despite some miscalibration. Such behavior also confounds the detection of miscalibration from the user's gaze behavior, as the system continues to receive inputs that align with its (miscalibrated) expectations.

In this work, we introduce a novel approach to seamlessly calibrate eye trackers. Our approach leverages a key novel insight about differences in gaze behavior during input (e.g. typing) versus output (e.g. reading): when providing system inputs through eye gaze (e.g. typing), users compensate for miscalibration to achieve desired system performance; in contrast, when users consume system outputs (e.g. by reading text they have typed) they do not compensate for miscalibration, thereby providing a signal for detecting the calibration offset (see Fig. \ref{fig:insight}). 
We demonstrate this approach through a novel autocalibrating gaze typing prototype \systemName{}, which tracks a user's reading behavior and compare their gaze to the location of typed characters on the screen to estimate the miscalibration amount and direction. \systemName{} aims to improve the gaze typing experience by seamlessly adjusting calibration in real-time, thereby reducing the need for manual recalibration and offering a more natural and efficient interaction. Our approach can potentially benefit a wide range of users, including those with motor impairments who use gaze keyboards for everyday communication and others who use gaze typing systems for extended periods, such as gamers and virtual/augmented reality headset users.

To explore the effectiveness of our approach, we used our \systemName{} prototype to run an in-lab user study with 20 participants and conduct a semi-structured interview with 6 ALS community stakeholders. 
Our user study results suggest that the proposed technique significantly reduces typing errors, improves typing speed, and enhances user satisfaction compared to a standard static calibration approach. Furthermore, feedback from our semi-structured interview suggests that the proposed technique can add value to users who frequently use gaze typing systems to communicate with their caregivers and the outside world, and shed light on desired future improvements.

In summary, our primary contributions are:
\begin{enumerate}
    \item The novel insight that gaze behavior differs when used for input (when miscalibration compensation may occur) vs. output (when compensation does not occur), and that this difference can be leveraged to seamlessly improve calibration.
    \item A novel gaze tracking autocalibration prototype called \systemName{}, which leverages this difference between eye gaze as input vs. output to seamlessly correct for miscalibration.
    \item An exploration of \systemName's effectiveness through a user study and semi-structured interview with stakeholders. Our results suggest that \systemName{} can significantly improve typing efficiency and experience, and shed light on needs it meets and desired future improvements.
\end{enumerate}

\section{Background and Related Work}
\label{sec:related}

We provide a brief overview of different eye movement categories identified in prior work (Sec~\ref{sec:eye_movement}), prior approaches for explicit and implicit calibration of eye trackers (Sec.~\ref{sec:rel_calibration}), and techniques beyond autocalibration that have been proposed in the past to make gaze typing interactions adaptive and seamless (Sec.~\ref{sec:rel_adaptive}).
Our work builds on this prior work, by learning from differences in eye gaze between input (e.g. typing) and output (e.g. reading) to improve seamless autocalibration.

\subsection{Eye Movement}
\label{sec:eye_movement}
Eye gaze movements can be characterized as: (a) fixations, (b) saccades, (c) smooth pursuits, and (d) vestibulo-ocular movements \citep{holmqvist2011eye}. Visual fixations maintain the focus of gaze on a single location. Fixation duration varies based on the task, but one fixation is typically 100-500 ms, although it can be as short as 30 ms. Saccades are rapid, ballistic, voluntary eye movements (usually between 20-200 ms) that abruptly change the point of fixation. Smooth pursuit movements are slower tracking movements of the eyes that keep a moving stimulus on the fovea. Such movements are voluntary in that the observer can choose to track a moving stimulus, but only highly trained people can make smooth pursuit movements without a target to follow. The gaze typing interface used in this work does not consist of any moving targets. Vestibulo-ocular movements stabilize the eyes relative to the external world to compensate for head movements. These reflex responses prevent visual images from slipping on the surface of the retina as head position changes. With respect to gaze typing on visual keyboards, we don't expect users to move their heads significantly. This is also valid for ALS users who progressively lose the ability to move their head with loss in muscle strength. Therefore, in our setup we expect users to primarily use fixations and saccades during gaze typing.

\subsection{Calibration Methods for Eye Trackers}
\label{sec:rel_calibration}
Calibration is a crucial aspect of eye tracking systems, 
mapping gaze points to locations in the field of view, requiring accurate data from eye trackers. However, it can be time-consuming, intrusive, and needs to be repeated frequently to maintain accuracy, as eye-tracking performance can degrade over time due to factors such as user movement or changes in ambient lighting \citep{jacob2003eye, cerrolaza2012error, wang2016deep, kasprowski2018comparison}.
Various calibration methods have 
been proposed to improve the accuracy and user experience of eye tracking devices. Traditional calibration techniques \citep{harezlak2014towards} involve the user following a series of on-screen targets, such as dots or markers (typically 9-point calibration), while the eye tracker records the user's gaze data (Fig. \ref{fig:calibration}). These methods, while effective, can be time-consuming and uncomfortable for some users, particularly those with motor impairments \citep{jacob2003eye, kasprowski2018comparison}.

\subsubsection{Explicit Gaze Calibration}
Prior works have proposed to reduce the time and effort required by the traditional 9-point calibration process through novel calibration interfaces where users are asked to explicitly fixate on fewer \citep{hoshino2020gaze} or more natural scene targets \citep{ohno2004free, saxena2022towards, kohlbecher2008calibration, vspakov2018enabling} to estimate parameters for a calibration function mapping the position of the eyes in 3D space to gaze coordinates on a 2D visual screen. 
\citet{cerrolaza2012error} estimate the calibration offset due to the user's head movement in the depth direction for a 4x4 grid on a 2D visual display. While these methods alleviate the burden of traditional calibration on an end-user, they all require the user's explicit cooperation with a novel calibration interface, and additionally might need custom hardware \citep{ohno2004free, kohlbecher2008calibration, vspakov2018enabling} and calibration interfaces \citep{saxena2022towards}. In contrast, our approach corrects for calibration errors without explicit cooperation of the user as they naturally perform the task at hand, without context switching to a new interface or additional custom hardware setups.

\subsubsection{Implicit Gaze Calibration}
Several prior approaches have also proposed implicit forms of eye gaze calibration (with no explicit cooperation from a user) either using saliency maps to determine natural target fixations in visual scenes \citep{kasprowski2018comparison, wang2016deep, hiroe2018implicit, hiroe2023implicit} or leverage assumptions that eye movements follow moving targets (smooth pursuits) \citep{pfeuffer2013pursuit, abdrabou2019calibration, bhatti2021eyelogin, lutz2015smoovs} and mouse clicks \citep{kasprowski2016implicit, gao2022x2t}.
\citet{sugano2015self} propose a self-calibrating approach for eye trackers based on a computational model of bottom-up visual saliency. This work assumes that the visual scene will have a user's gaze fixations always lie inside a small cluster of salient regions in the egocentric view of the user. While this approach is implicitly adaptive and leverages natural junctures of the user's visual view on a screen or the 3D environment, it is data intensive for accurate autocalibration. Similarly, \citet{kasprowski2018comparison} propose a calibration technique for headset eye trackers that maximizes the likelihood of 2D gaze coordinates falling in regions of high saliency. In comparison, our approach relies on natural elements of the user interface as fixation targets, leverages the difference between gaze behaviors as an input versus an output modality, and is sample efficient (autocalibrating effectively with fixation on one target point).

\citet{gao2022x2t} present an adaptive learning approach which autocalibrates an RGB-based eye tracker for a custom designed circular keyboard (words to be typed next are displayed on the circle). However, they assume the user can provide click-based feedback to the learning system in the form of backspaces activated on a physical keyboard/device. In comparison, our method only relies on feedback from the gaze trajectory available on the visual keyboard and does not use any external hardware other than the eye tracker.

\citet{lutz2015smoovs} designed a custom gaze typing interface (clusters of characters placed in a circle which move outwards) for calibration-free text entry on public displays by learning correlations between eye movements and target movements. Similarly, \citet{bhatti2021eyelogin} and \citet{abdrabou2019calibration} use smooth pursuit eye movement correlations with visual key movements for calibration-free text entry systems.
While these approaches alleviate the need for explicit calibration prior to gaze typing, they rely on custom designed visual keyboards with moving targets for implicit calibration. Our proposed approach instead is flexible and can augment any existing keyboard layout and eye tracker, relies on gaze fixation during natural junctures of the typing task while leveraging  differences between gaze behaviors during perception versus control.

\subsection{Other Adaptive Gaze Typing Techniques}
\label{sec:rel_adaptive}
Adaptive techniques for gaze typing aim to improve the efficiency and user experience of gaze-based text entry by dynamically adjusting system parameters based on the user's performance and gaze behavior. 
To enable widespread adoption of gaze typing technologies for AR applications, \citet{schenkluhn2022look} emphasize the need to create gaze typing that proactively adapts dwell time instead of retrospectively reacting to user fatigue. This would enable users to type short texts at their peak performance and economically utilizing cognitive resources for long texts. \citet{mott2017cascading} proposed a cascading dwell technique that automatically adjusts the dwell time for gaze-based input based on the user's performance. This approach has been shown to improve typing speeds and reduce errors in text entry tasks, highlighting the importance of dynamic adjustments in gaze-based input systems. \citet{cui2023glancewriter} propose to alleviate the effort required to type with dwell-based gaze typing systems by probabilistically predicting next characters to type from natural glances at visual keys without any need to dwell or specify the starting and ending positions of a gaze path when typing a word. Our autocalibration approach can augment such dwell-free gaze typing systems where differences in gaze behaviors during reading and typing still persist.

\citet{chen2021adaptive} propose an adaptive gaze typing system using a computational model of the control of eye movements in gaze-based selection. They formulate the model as an optimal sequential planning problem bounded by the limits of the human visual and motor systems and use reinforcement learning to approximate optimal solutions for number of fixations and duration required to make a gaze-based selection. While these approaches present adaptive learning based solutions for dwell-time customization, we propose a technique which adaptively corrects for miscalibration of an eye tracker during gaze typing that can be combined with dwell-time customization to further enhance user experience in the future. 
\section{E\lowercase{ye}O Prototype} 
\label{sec:algo}

Current eye-gaze systems enable users to stop what they are doing and manually recalibrate if they think their calibration is off. We present a prototype, \systemName{}, that automatically recalibrates \textit{while the person is typing} without the need for manual recalibration. 
Our approach is based on the differences in gaze behavior during input (perception) and output (control). When an eye tracker's miscalibration is detected by a user during the typing task, they may attempt to compensate for the offset in the detected gaze coordinates to type the characters of interest. 
However, when the user looks up at the text box, and fixates to read the text they have typed, they do not need to compensate for miscalibration since they are not attempting to control the system. In this way, reading provides a free calibration signal, as eye gaze is not being compensated for miscalibration (but may occur during typing). By monitoring the difference in users' eye movements during typing versus reading, our technique dynamically adjusts the calibration in real-time, compensating for miscalibration errors.

\subsection{System Design}
\label{sec:interface}

\begin{figure*}[ht]
  \centering
  \subfigure[]{
  \includegraphics[width=0.40\linewidth]{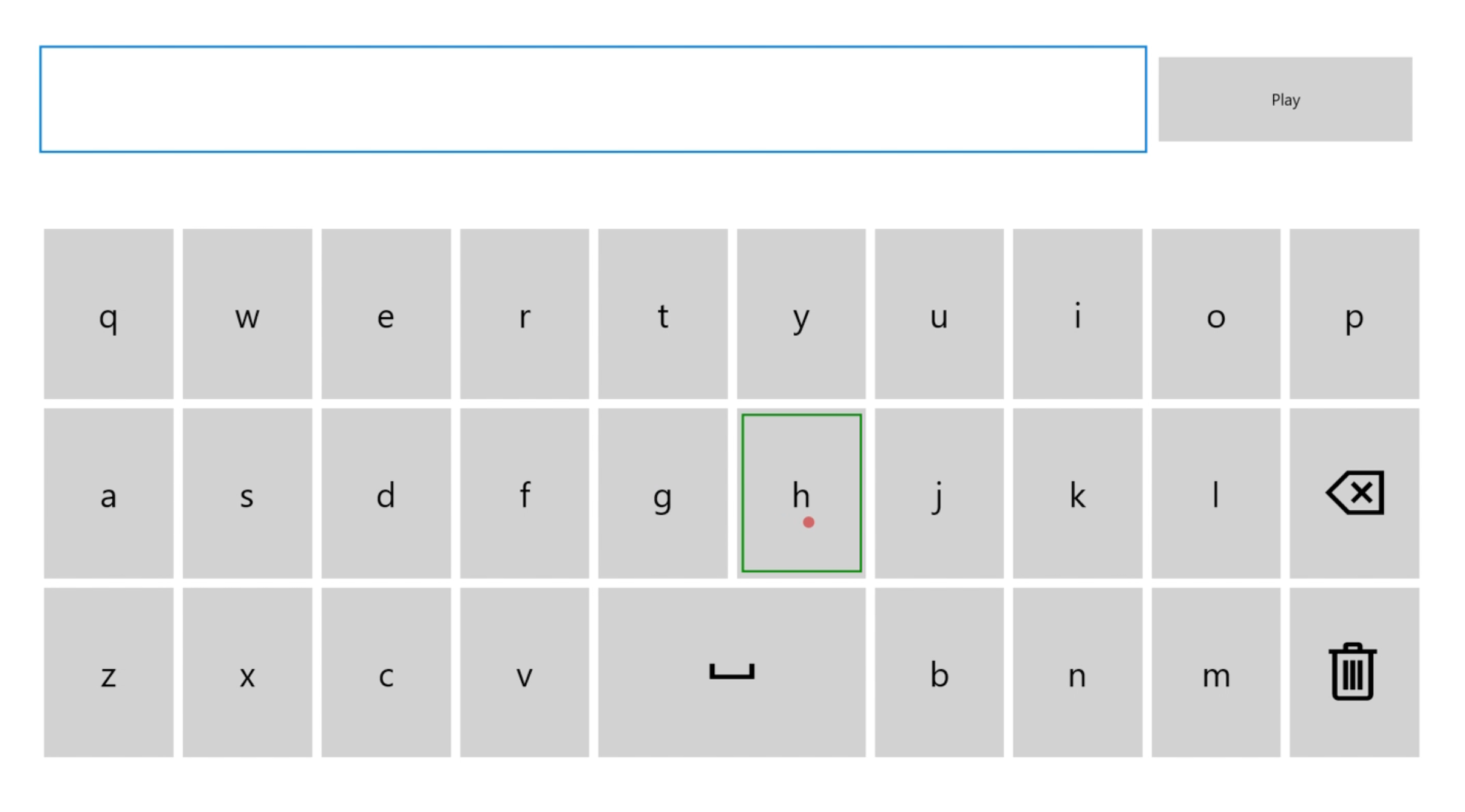}
  }
  \hspace{1mm}
  \subfigure[]{
  \includegraphics[width=0.40\linewidth]{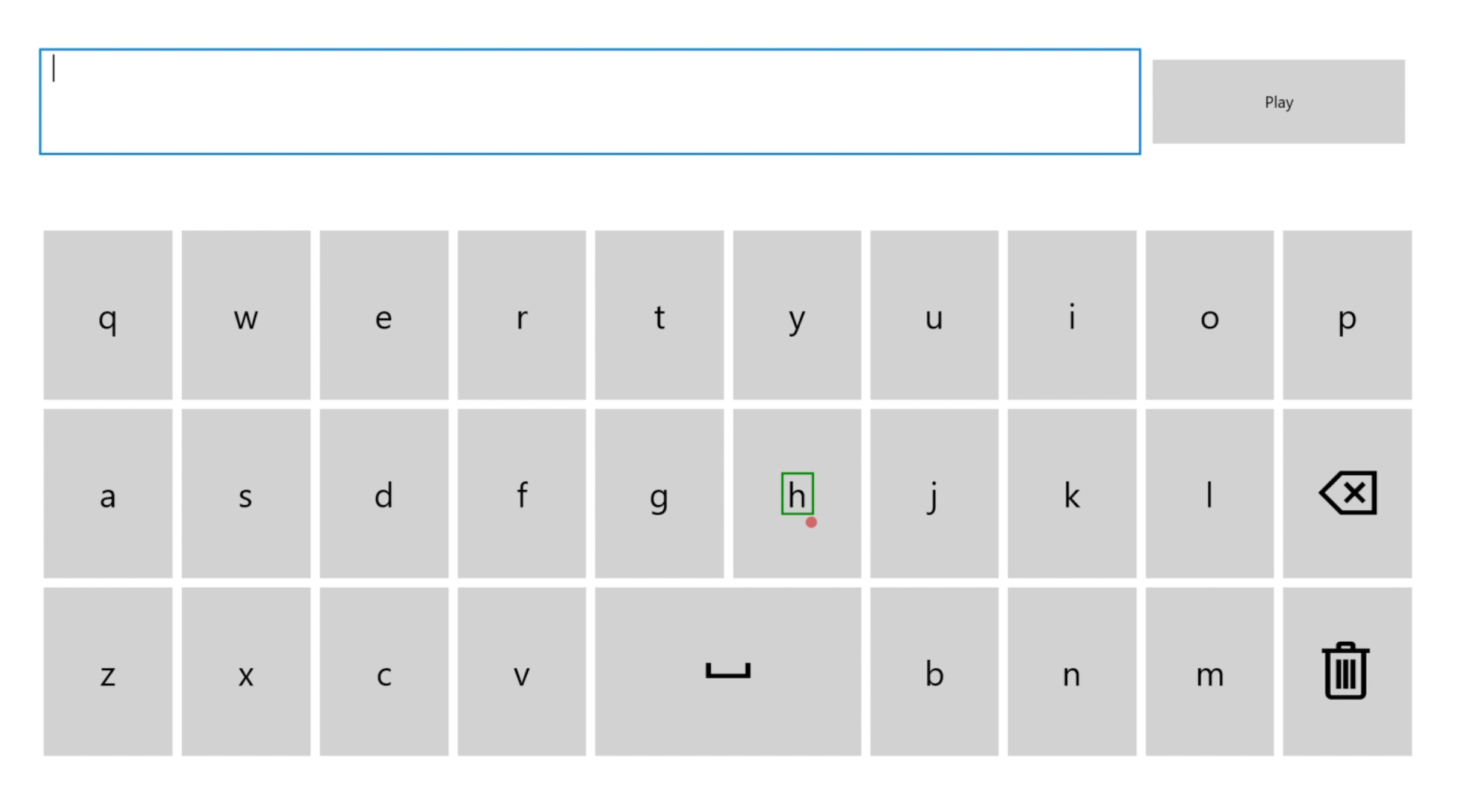}
  }
  \hspace{1mm}
  \subfigure[]{
  \includegraphics[width=0.40\linewidth]{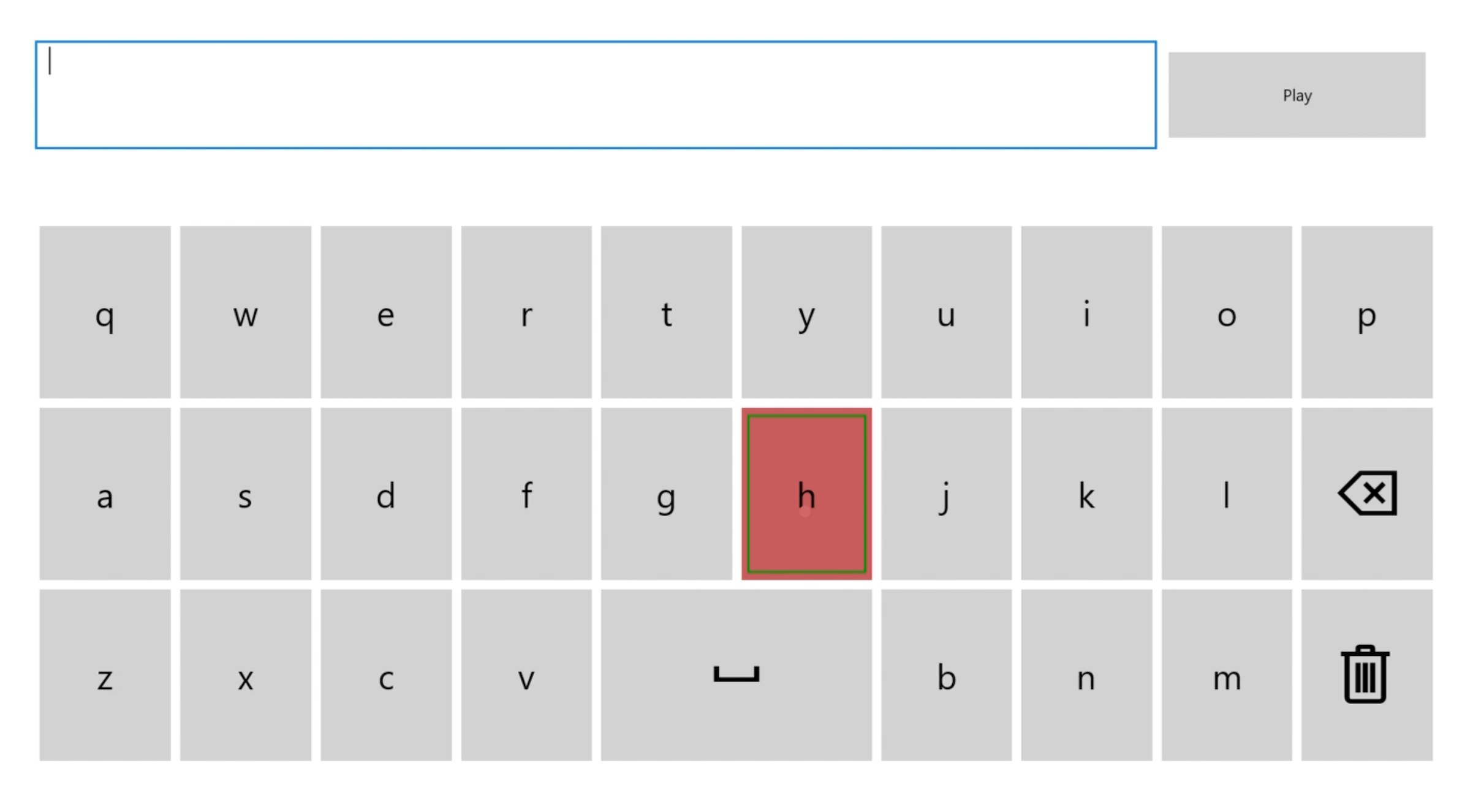}
  }
  \subfigure[]{
  \includegraphics[width=0.40\linewidth]{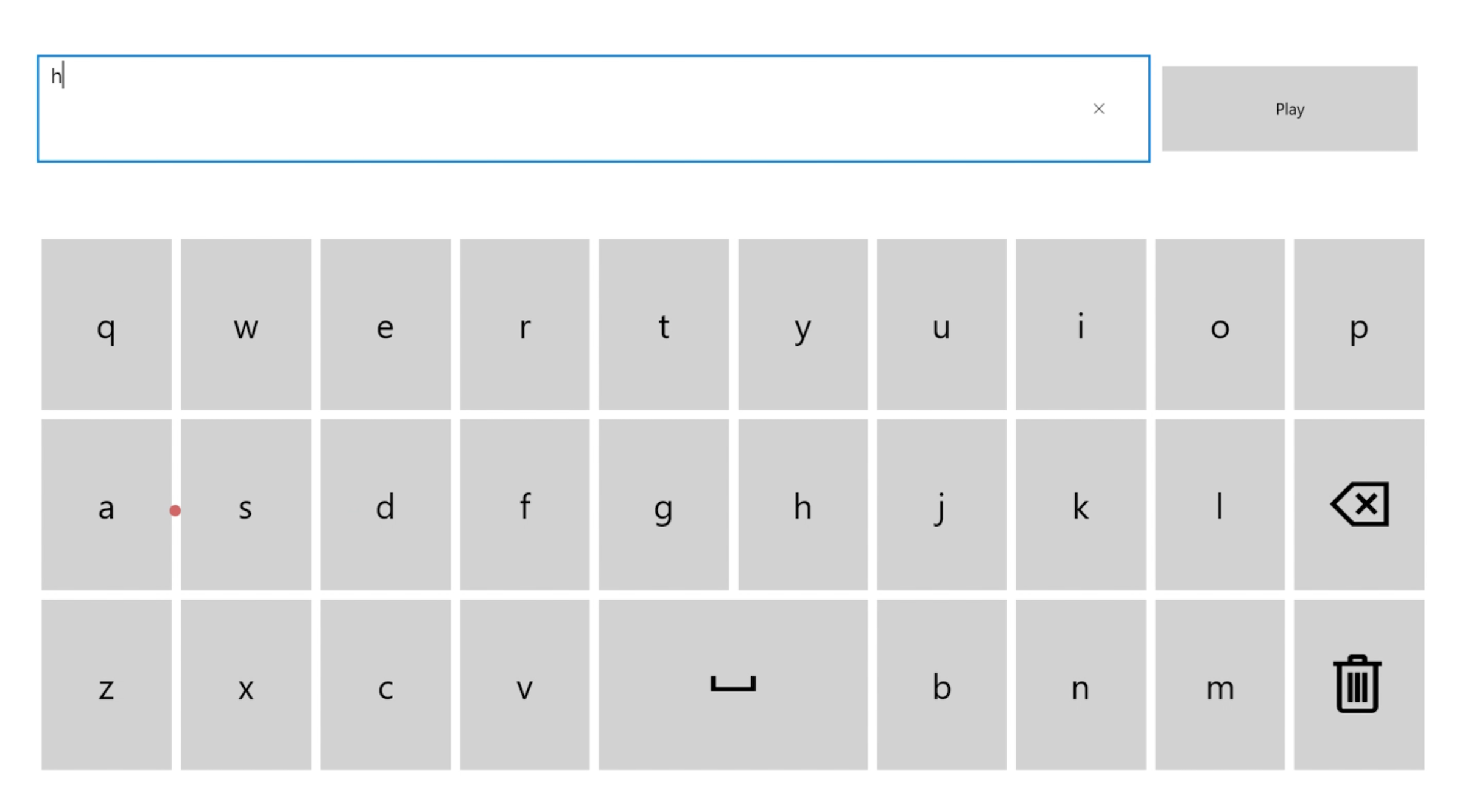}
  }
  \caption{Screenshots of the system's visual display. A user can activate a key (i.e. type a character) on the visual keyboard by dwelling on it for a fixed duration of time. Here we show snippets of visual feedback (in order) that the user receives when successfully activating the key for the letter `h'.}
  \label{fig:activation}
\end{figure*}

\systemName{} is functionally similar to currently available visual keyboards, such as in Windows Eye Control \citep{wec}. However, it provides more control to process and update the miscalibrated gaze coordinates in real-time. Figure \ref{fig:activation} illustrates the mechanism for typing a character on the visual keyboard through eye gaze. A user's detected gaze location is displayed as a red dot on the screen. If they fixate on a key for $50$ ms, a dwell timer of $400$ms is initiated by the system. The start of the timer is depicted by a green rectangle on the selected key (Fig. \ref{fig:activation}(a)). During fixation on the key, the rectangle slowly decreases in area, eventually collapsing at the center of the key (Fig. \ref{fig:activation}(b)). When the timer finishes, the user receives visual feedback that the character is typed: the key turns red and the letter is added to the text box at the top (Fig.~\ref{fig:activation}(c)). 

\begin{figure*}[ht]
  \centering
  \subfigure[Windows Eye Control application]{
  \includegraphics[width=0.35\linewidth]{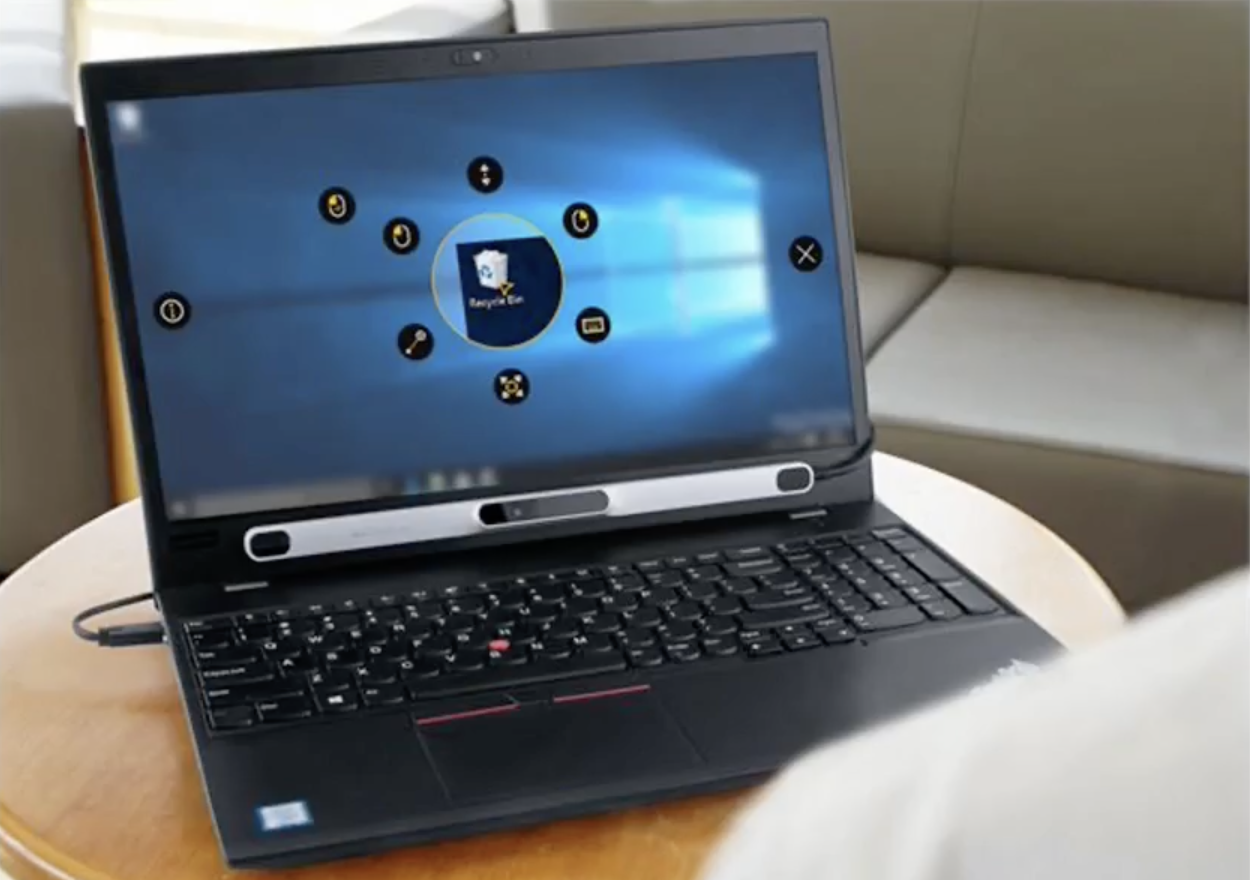}
  }
  \subfigure[A user operating a customized gaze typing application]{
  \includegraphics[width=0.37\linewidth]{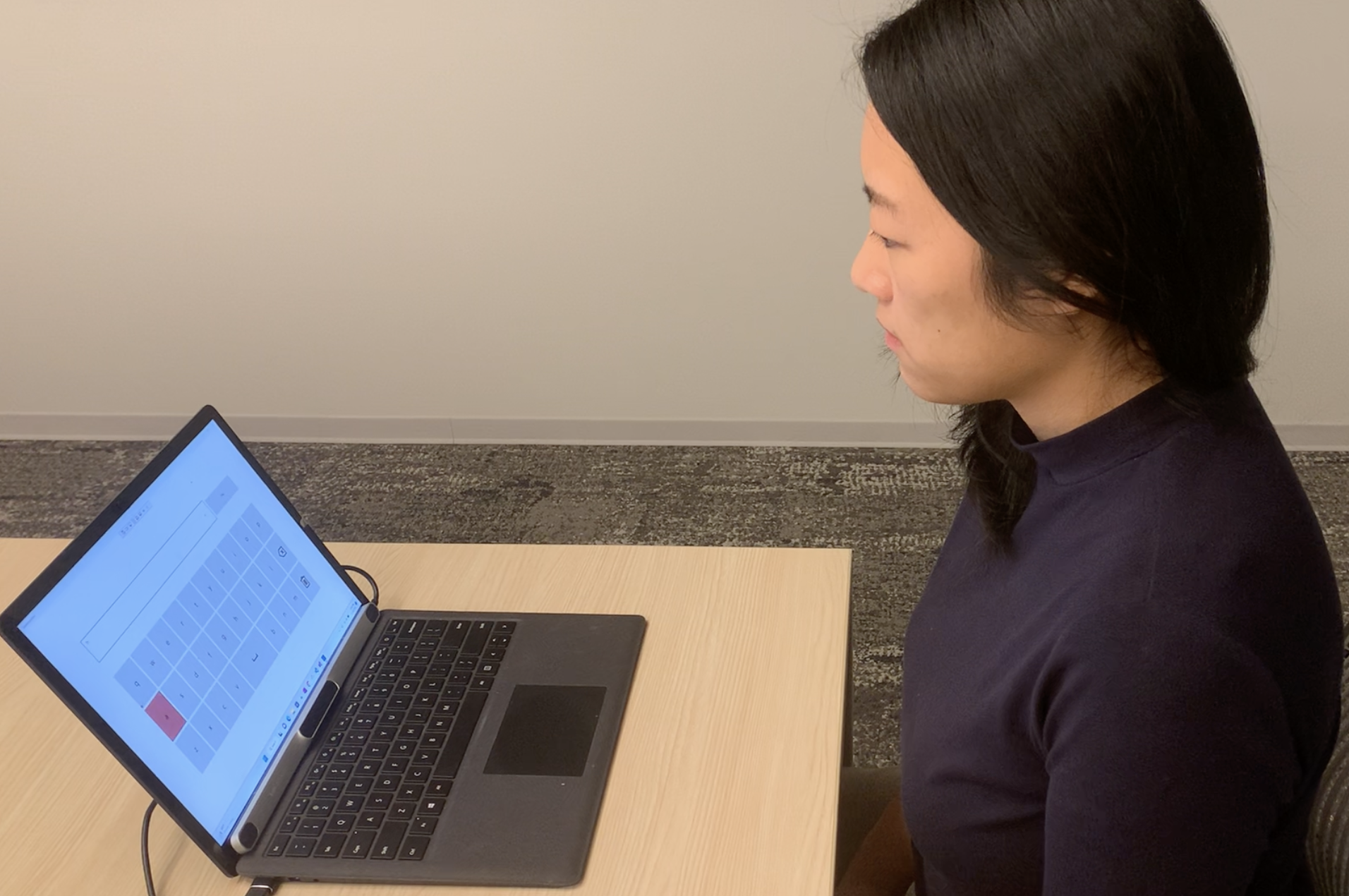}
  }
  \caption{Windows applications for (a) visual PC control and (b) gaze typing respectively which leverage a user's eye movements tracked via an external Tobii eye tracking device.}
  \label{fig:gazetyping}
\end{figure*}

Our system utilized a infrared Tobii PCEye eye tracker \citep{tobii}, with a sampling rate of 60 Hz, connected to a standard Windows 11 laptop via USB (Fig.~\ref{fig:gazetyping}). The eye tracker can be calibrated via a standard Tobii software available with the purchase of the tracker (Fig.\ref{fig:calibration}). It requires a user to follow a set of dots (in sequence) on the screen to calibrate their gaze. Users receive feedback about how the calibration process went and if satisfied with the results, they can then return to interacting with any application of interest.

We work with a customized gaze typing application. The tracked gaze is directed towards an on-screen visual keyboard (Fig. \ref{fig:activation}), displayed on a 24-inch laptop screen. 
The on-screen keyboard came as part of an UWP application which was customized on top of the Microsoft Gaze Interaction Library \citep{gil}. The application was written using the Tobii Pro SDK (designed to offer access to gaze data from Tobii eye trackers) in C\#. It facilitates the capture and processing of x, y coordinates from the eye tracker, providing necessary data for \systemName{} to leverage and update the gaze coordinates in real-time. 
The software for our autocalibrated gaze typing technique was developed in Python, which received the 2D gaze coordinates in real-time from the UWP application via Google's remote procedure call (RPC) protocol. This allowed Python scripts to access gaze data, analyze it, and apply the necessary miscalibration corrections to display back on the visual keyboard application.

\subsection{Gaze Filtering}
In our work, we primarily distinguish between fixations and saccades (see Section \ref{sec:related}). We assume smooth pursuit movements are not present in our trials as the visual keyboard has no moving targets. To detect fixations, we consider spatial and temporal criteria from the taxonomy of fixation identification algorithms described by \citet{salvucci2000identifying}. We use velocity-based criteria from spatial characteristics and duration based criteria from temporal characteristics, to filter out fixations from saccades. 
\citet{salvucci2000identifying} proposed a novel taxonomy of fixation identification algorithm and evaluated existing algorithms in the context of this taxonomy. They identify two characteristics---spatial and temporal---to classify different algorithms for fixation identification. For spatial characteristics, three criteria distinguish primary types of algorithms: velocity-based, dispersion-based, and area-based. For temporal characteristics, they include two criteria: whether the algorithm uses duration information, and whether the algorithm is locally adaptive. The use of duration information is guided by the fact that fixations are rarely less than 100 ms and often in the range of 100-500 ms. In our work, we use velocity-based and area-based criteria under spatial characteristics and duration-based criteria under temporal characteristics to filter out fixations from saccades. We first filter out eye movements with very high speeds (a large distance traversed over a very short period of time is likely a saccade). If gaze
continues to not be classified as a saccade for more than 100 ms, then we consider it a fixation.

\subsection{Autocalibration Algorithm}

\begin{figure}
    \centering
    \includegraphics[width=0.95\linewidth]{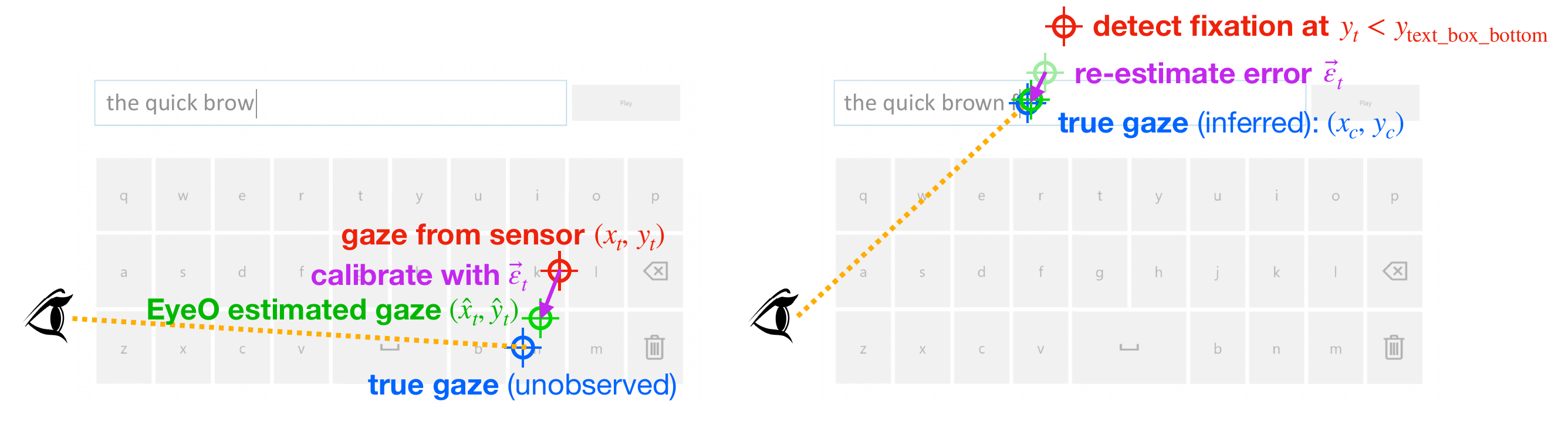}
    \caption{Simplified diagram of Algorithm~\ref{alg:main}, depicting the input and output modes of the EyeO auto-calibrating interface. \emph{Left:} While the user is typing (lines 15-22 \& Figure~\ref{fig:gazetyping}a), the algorithm receives a stream of miscalibrated sensor inputs $(x_t, y_t)$, and sends a calibrated sequence $(\widehat x_t, \widehat y_t) := (x_t + \epsilon_{x_t}, y_t + \epsilon_{y_t})$ to the interface, adjusted using the current error estimates. \emph{Right:} While the user is consuming system output (lines 8-13 \& Figure~\ref{fig:gazetyping}b), the system infers that the user is looking at $(x_c, y_c)$, the location of the last typed character, and uses this to \emph{update} the error estimates $(\epsilon_{x_t}, \epsilon_{y_t})$ used for calibration. This inference only occurs when fixation is detected within a threshold $\tau$.} 
    \label{fig:alg}
\end{figure}

In this section, we describe the low-level autocalibration algorithm implemented in EyeO. Figure~\ref{fig:alg} provides a simplified schematic of the procedure, while Algorithm~\ref{alg:main} presents it in full detail. The algorithm receives a stream of coordinates $(x_t, y_t)$ from the device driver, as well as several variables describing the state of the keyboard interface. It produces estimated gaze coordinates $(\widehat x_t, \widehat y_t)$ based on adaptively estimated calibration errors $(\epsilon_{x_t}, \epsilon_{y_t})$, which are refined whenever the user is inferred to be ingesting the system's output (i.e. reading from the text box).

The user's gaze is assumed to be directed towards the text box when it falls above the keyboard layout and within a certain threshold distance $\tau$ from the center of the last typed character. 
If the gaze is detected within this area (calibration zone), the system then identifies whether the user's gaze is in a state of fixation or saccade. Visual fixations maintain the focus of gaze on a single location. 
We first filter out eye movements with very high speeds (a large distance traversed over a very short period of time is likely a saccade), and then detect fixations within the zone of calibration. If gaze continues to not be classified as a saccade for more than 100 ms and lies within the calibration zone, we consider it a fixation for reading.

\begin{algorithm}
\caption{EyeO Autocalibration}
\begin{algorithmic}[1]
\renewcommand{\COMMENT}[1]{\hfill\textcolor{gray}{// #1}}
\REQUIRE{Stream of gaze coordinates $(x_t, y_t)$ detected by an eye tracker; window size $w$ for running average; detected calibration error bound $b$; calibration zone threshold $\tau$; $y$-coordinate for lower boundary of text box $y_\mathrm{text\_box\_bottom}$}
\STATE Initialize $\tau := \tau_\mathrm{init}$ 
\COMMENT{threshold pixel distance of gaze coordinates from last typed character}\\
\STATE Initialize $n_\mathrm{char} := 0$ \COMMENT{number of characters visible in text box}\\
\STATE Initialize $\epsilon_{x_0} := 0, \epsilon_{y_0} := 0$ \COMMENT{calibration error in $x$ and $y$ directions}\\
\FOR{raw gaze coordinates $(x_t, y_t)$:}
\STATE Receive $n_\mathrm{char} 
$from typing application
\IF{$y < y_\mathrm{text\_box\_bottom}$  \textbf{and} $n_\mathrm{char} > 0$}
\STATE{Receive location of last type character ($x_c, y_c$) from typing application}
\IF{fixation detected \textbf{and} $\sqrt{(x_c-x_t)^2 + (y_c-y_t)^2} < \tau$}
\STATE{$\delta_{x_t} := x_c-x_t, \quad \delta_{y_t} := y_c-y_t$} \COMMENT{infer user is reading at $(x_c, y_c)$}\\
\STATE{$\delta_{x_t} := \frac{1}{\min(w,t)} \sum_{j=\max(0,t-w+1)}^{t} \delta_{x_t}$, \quad $\delta_{y_t} := \frac{1}{\min(w,t)} \sum_{j=\max(0,t-w+1)}^{t} \delta_{y_t}$} \COMMENT{sliding window estimate}\\
\STATE{$\epsilon_{x_t} := \mathrm{clip}(\delta_{x_t}, -b, b), \quad \epsilon_{y_t} := \mathrm{clip}(\delta_{y_t}, -b, b)$} \COMMENT{clip to maximum allowed offset}\\
\ENDIF
\ELSE
\STATE{$\epsilon_{x_t} := \epsilon_{x_{t-1}}, \quad \epsilon_{y_t} = \epsilon_{y_{t-1}}$} \COMMENT{infer user is typing; keep current error calibration}\\
\ENDIF
\STATE \textbf{return} calibrated gaze coordinates $(\widehat x_t, \widehat y_t) := (x_t+\epsilon_{x_t} + y_t+\epsilon_{y_t})$
\ENDFOR
\end{algorithmic}
\label{alg:main}
\end{algorithm}

We assume that the user reads the last typed character when they look up to read and detect the offset in calibration accordingly. 
Our system computes a moving average of the calibration offsets, 
which enables the calibration to be updated continuously and smoothly in real-time. 
The system continuously updates the calibration error based on the user's gaze behavior while reading the typed text. The computed correction to the detected miscalibration is continuously applied when the user's gaze moves away from the text box (i.e. while typing on the visual keyboard). The autocalibrated gaze coordinates are displayed to the user with the updated location of the red gaze cursor on the screen (Fig. \ref{fig:activation}(a)). 
The approach is detailed in Algorithm \ref{alg:main}.
We instantiate this algorithm with calibration zone size parameter
$\tau = 150$,
window size $w=64$, and calibration error bound $b=200$. 
\section{User Study}
\label{sec:user-study}

To explore the effectiveness of our autocalibration technique and how it shapes a user's gaze typing experience, we ran an in-lab user study with IRB approval. We compared \systemName{} to a standard manual calibration control. 
After standard calibration, miscalibration were purposefully introduced to evaluate the typing experience of the user under such conditions.
Participants typed a set of 5 unique phrases each for different miscalibration  under the two gaze typing systems. The users were made aware that the system will have miscalibrations during the study, but were not made aware of any differences between the two systems. 

\subsection{Participants}

We recruited 20 participants (14 males, 6 females) aged 18-55 who were members of organizations the authors of this work were affiliated with. All participants sighted (i.e. they did not need glasses, or their vision was correctable with glasses) and did not report any eye conditions. Two wore corrective eye glasses during their respective session. Each session lasted around 60 minutes and the participants were reimbursed through purchased gifts. Two participants had prior experience in gaze typing. One participant found the typing experience very uncomfortable during the practice session and opted out of the study. We thus report results for 19 participants.

\subsection{System Setup}
\begin{figure*}[ht]
  \centering
  \includegraphics[width=0.65\linewidth]{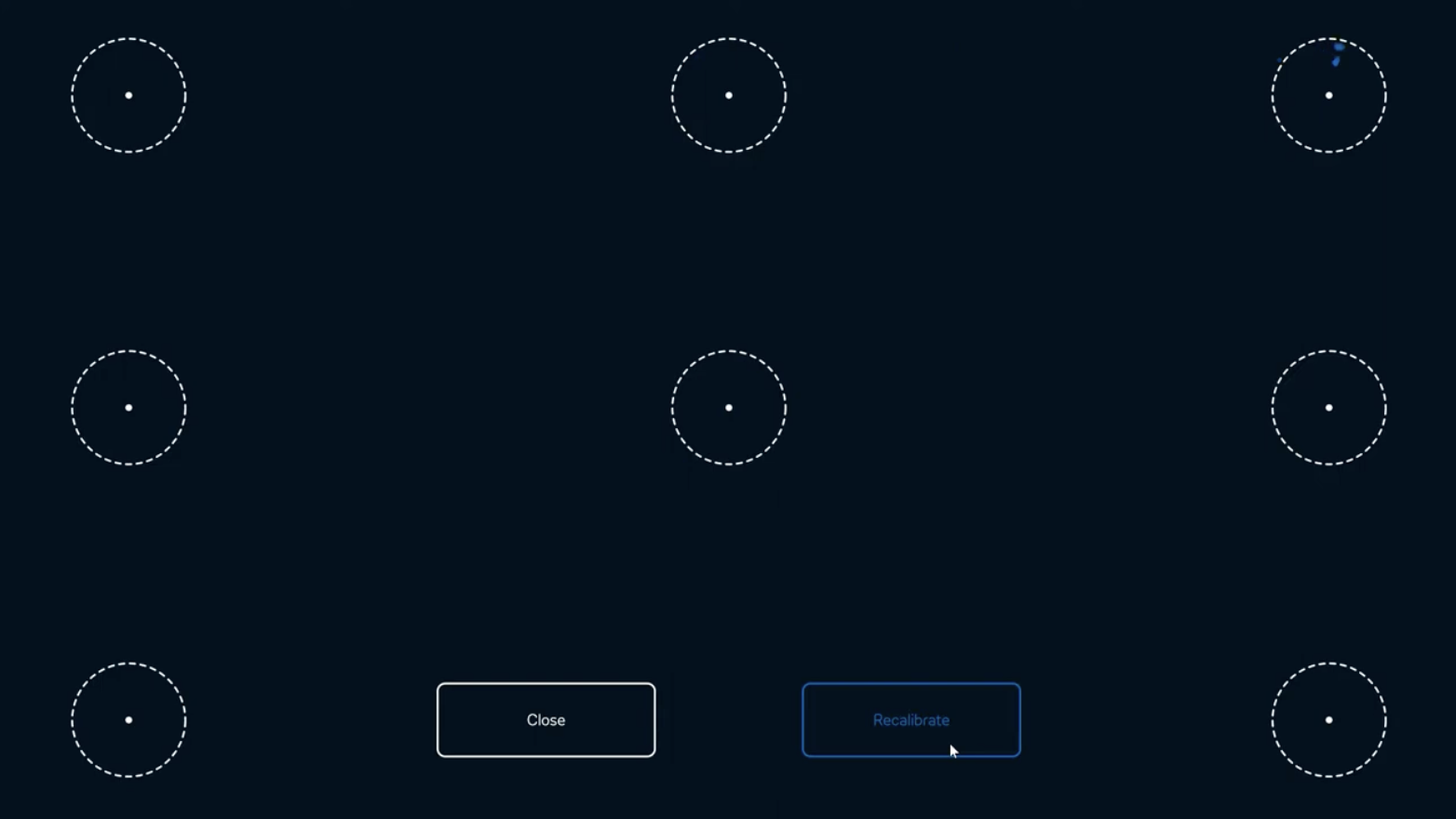}
  \caption{Traditional 9-point calibration system used to calibrate eye tracking hardware.}
  \label{fig:calibration}
\end{figure*}

Participants were seated comfortably at a distance of approximately $75$cm from a computer monitor and head movements were not restrained. Each participant engaged with two gaze typing systems: 1) \systemName, which leverages differences between reading and typing to autocalibrate, and 2) a control that relied on traditional calibration alone. Both systems shared the exact same gaze typing interface. The only difference was that \systemName{} contained an additional layer of Python software to detect and correct for miscalibration (described in Section \ref{sec:algo}). 
We used a desktop-based tracker (typically attached to the bottom of a computer screen), and in particular a head-pose free tracker, that let the user move their head within a certain range of locations and orientations. All manual calibrations were performed via the standard Tobii SDK 9-point calibration procedure (Fig. \ref{fig:calibration}).

\subsection{Procedure}
Participants were initially introduced to gaze typing and given a brief tutorial on how to use the on-screen keyboard. The experiment consisted of an initial calibration procedure, a practice round, and trials of both \systemName{} and the control (within-subjects study). The order of the \systemName{} and control phases was counterbalanced across participants. After using each system, users provided feedback about their experience with the system via the NASA TLX questionnaire \citep{hart1988development}. Users were free to abort a typing session if they found the setup too cumbersome to continue typing.

\subsubsection{Initial Calibration and Practice Session}
All participants first performed a manual calibration. Next, participants completed a practice round in which they typed a couple of simple phrases (`happy new year', `hello world') in the gaze typing application.

\subsubsection{System Trial (x2)}
\label{sec:trial}
For each system (\systemName{} and Control), participants first completed a manual calibration and were then asked to type 5 different phrases (1 per session). 
At the start of each of the 5 sessions, we altered the calibration of the tracker by a fixed amount of 0 (no miscalibration) or +75, -75 pixels in the x-direction, or +75, -75 pixels in the y-direction (counterbalanced order between sessions). Errors were introduced independently in the x and y directions to better isolate the impact of compensation in the two directions. This design decision was made because we learned via a pilot study that some users are more comfortable compensating in one direction versus the other. Additionally, we also learned from the pilot study the average amount of miscalibration users could work with (by compensating). The induced miscalibrations affected the entire screen uniformly.  

\subsubsection{Overall Preferences} 
\label{sec:interview}
To better understand if \systemName{} led to a more positive user experience over the control system, we asked participants to rate each system in terms of the mental workload required (right after typing 5 phrases with it) using the NASA TLX questionnaire \citep{hart1988development}. At the end of the study, we also asked them to share their opinions in written form about how they compared their experience of gaze typing between the two systems, and if they preferred one system over the other.

\subsection{Typed Phrases}
\label{subsec:phrases}

The phrases that participants typed were randomly selected from a standard corpus to evaluate text entry systems \citep{mackenzie2002text}, representing a diverse range of words and character combinations. No two phrases were repeated across sessions for the same participant or across sessions for different participants.

\subsection{Metrics}
\label{sec:metrics}
To assess the effectiveness of \systemName{} autocalibration to ease the task of gaze typing and it's impact on user experience, we compared typing efficiency in terms of typing speed (characters per minute), number of backspaces, and abort frequency (number of sessions where the user gave up typing the phrase) as quantitative metrics. For quantitative subjective feedback, we compare NASA TLX \citep{hart1988development} responses and user preferences between the control and the \systemName{} typing systems. A two-sided t-test was performed to determine the significance of any observed differences. We also study qualitative subjective feedback from open-ended responses.
\subsection{Results}
\label{sec:results}

\subsubsection{Typing Efficiency}

\begin{figure*}[ht]
  \centering
  \subfigure[Typing Speed]{
  \includegraphics[width=0.22\linewidth]{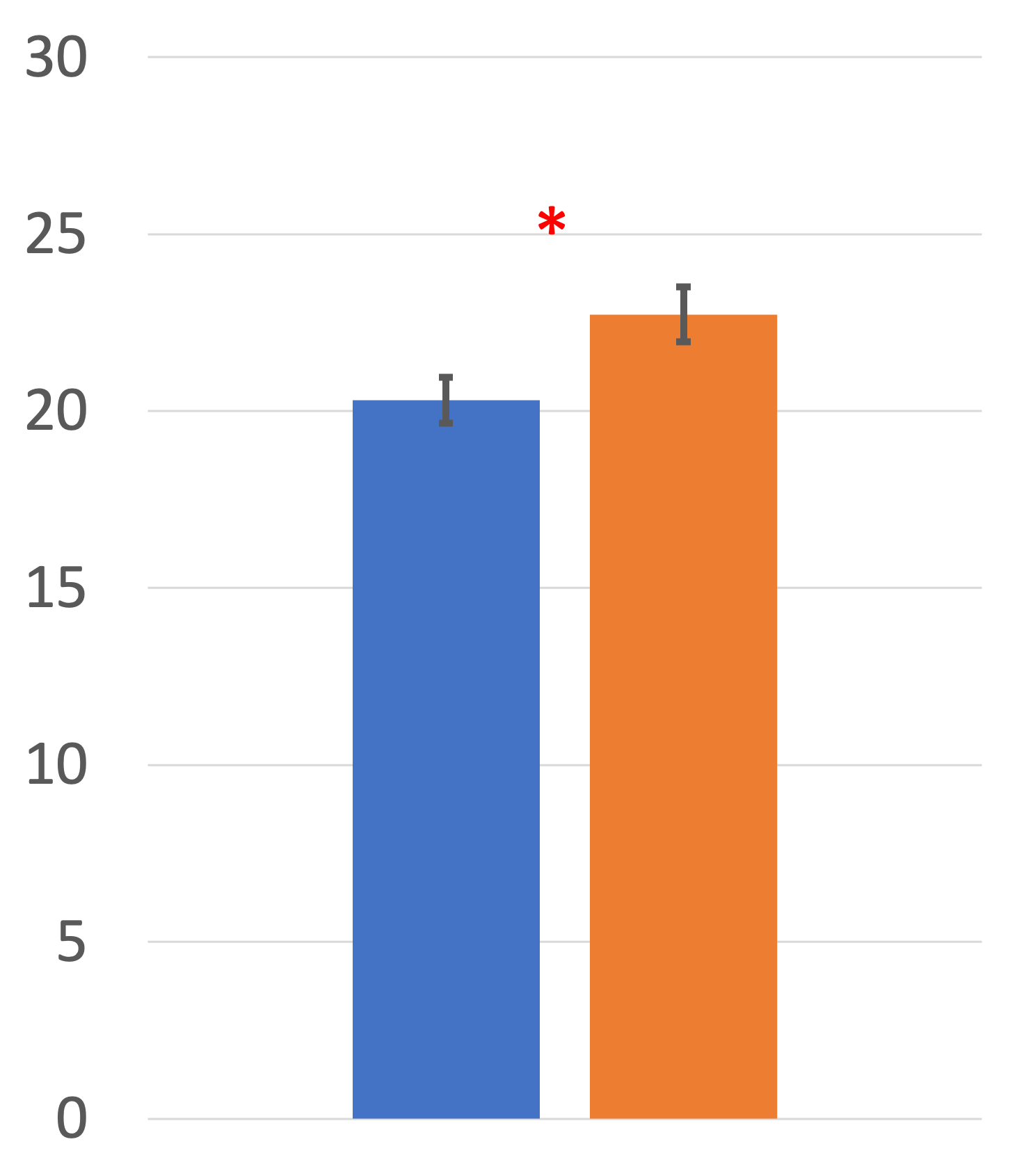}
  }
  \subfigure[Abort Frequency]{
  \includegraphics[width=0.22\linewidth]{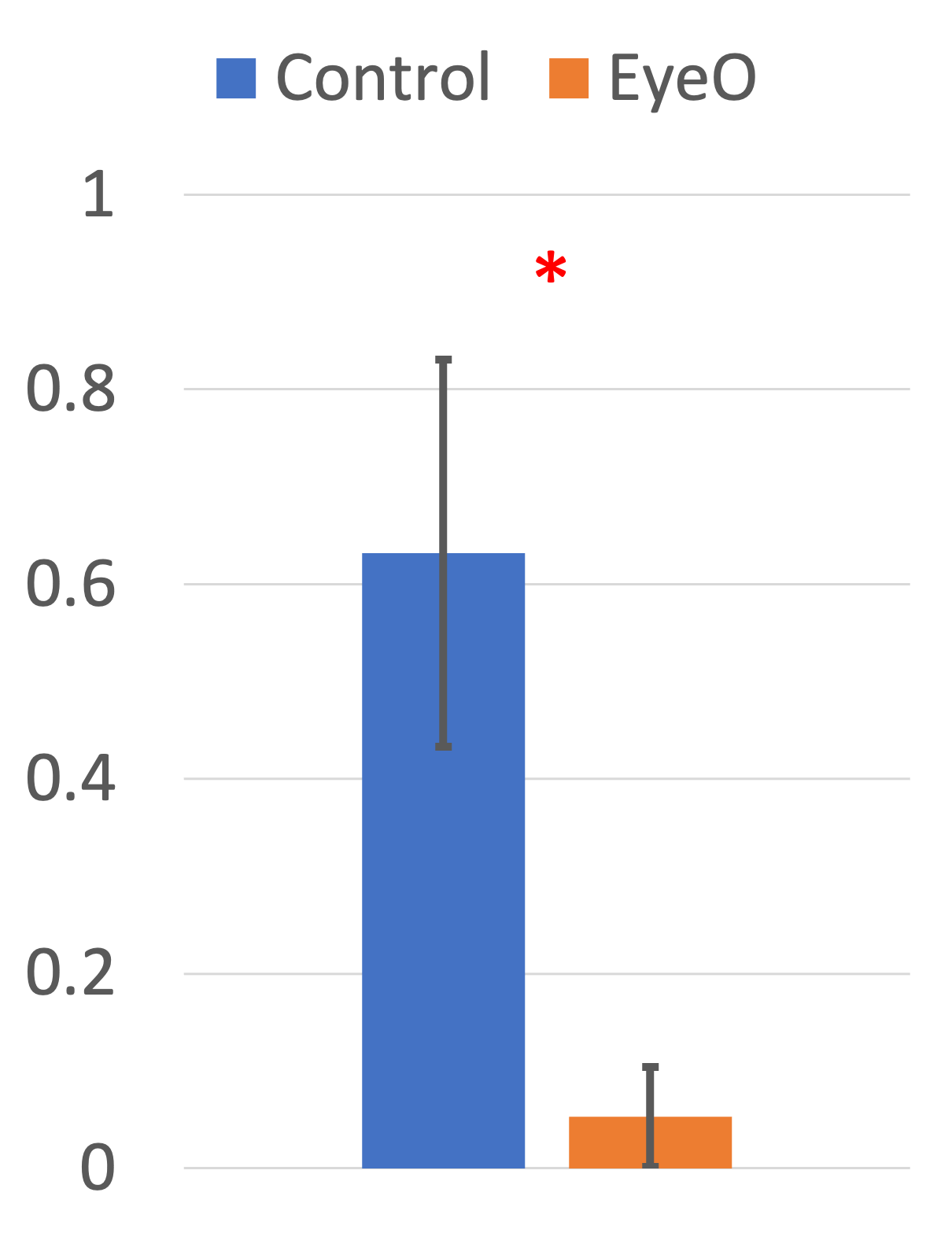}
  }
  \subfigure[Number of Backspaces]{
  \includegraphics[width=0.20\linewidth]{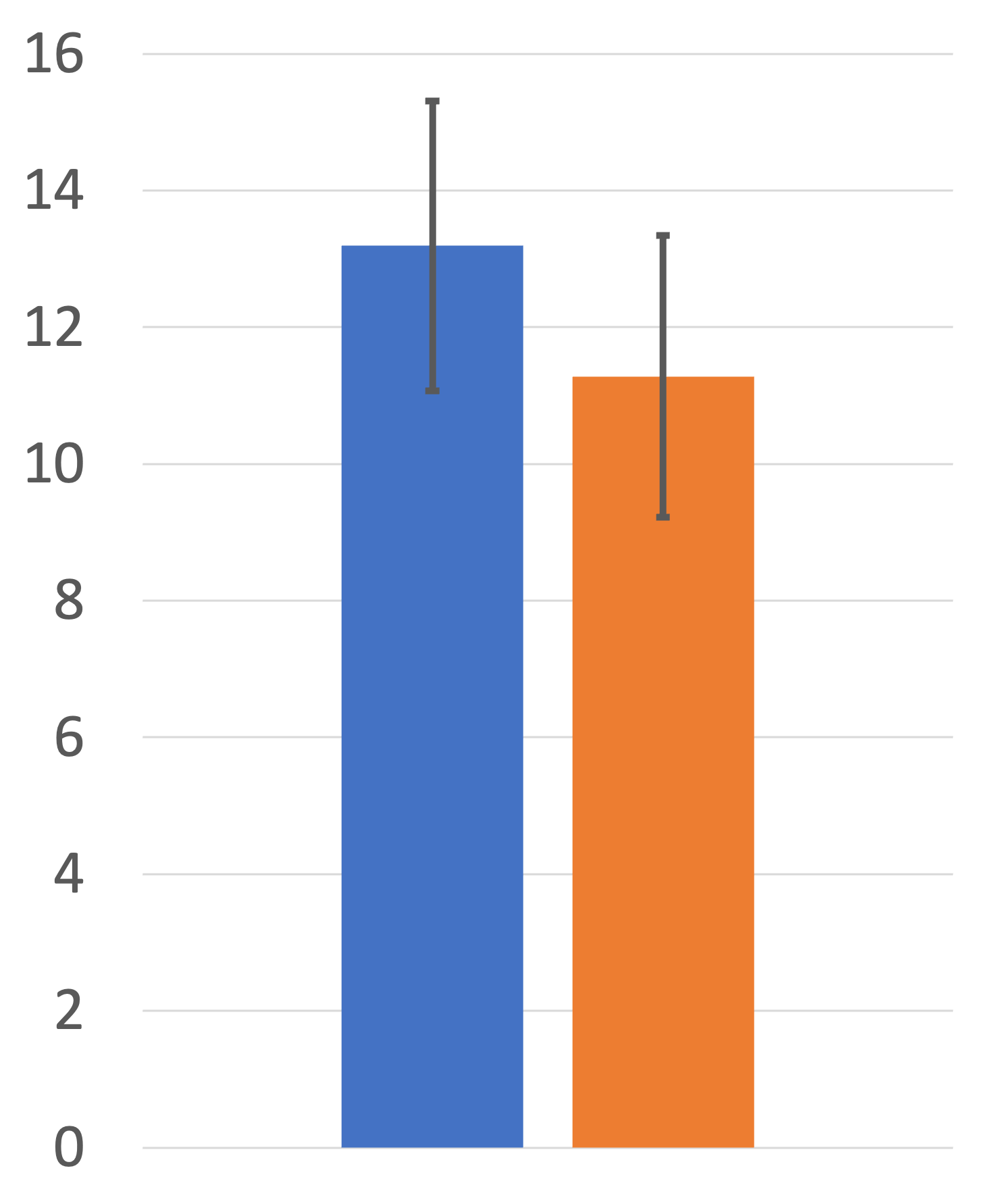}
  }
  \caption{Quantitative metrics of typing efficiency for the two systems. Significance codes: * < .05, ** < .01, *** <.001}
  \label{fig:quant}
\end{figure*}

We found that \systemName{} exhibited faster typing speeds (characters/minute) (Fig. \ref{fig:quant}(a); $p<0.05$), lower abort frequency (Fig. \ref{fig:quant}(b); $p<0.05$), and required fewer backspaces (Fig. \ref{fig:quant}(c); $p=0.53$) in comparison to the static control system. Differences between typing speed (\systemName{}: M=$22.73$, SE=$0.78$; Control: M=$20.31$, SE=$0.65$) and abort frequency (\systemName{}: M=$0.05$, SE=$0.05$; Control: M=$0.6$, SE=$0.19$) were statistically significant (Fig.~\ref{fig:quant}(a),(c)). 
When we consider the gaze data for \systemName{} after the first reading attempt (when the system gets the first opportunity to autocalibrate), differences for typing speed 
(\systemName{}: M=$25.91$, SE=$0.98$; Control: M=$20.31$, SE=$0.65$; $p<0.001$) and number of backspaces used 
(\systemName{}: M=$9.13$, SE=$2.0$; Control: M=$13.19$, SE=$2.12$; $p=0.18$) were even larger. With this consideration, typing speed is  strikingly different between the two systems ($p<0.001$), hinting towards the effectiveness of \systemName{}'s autocalibration technique.

\subsubsection{Mental Workload}

\begin{figure*}[ht]
  \centering
  \includegraphics[width=0.8\linewidth]{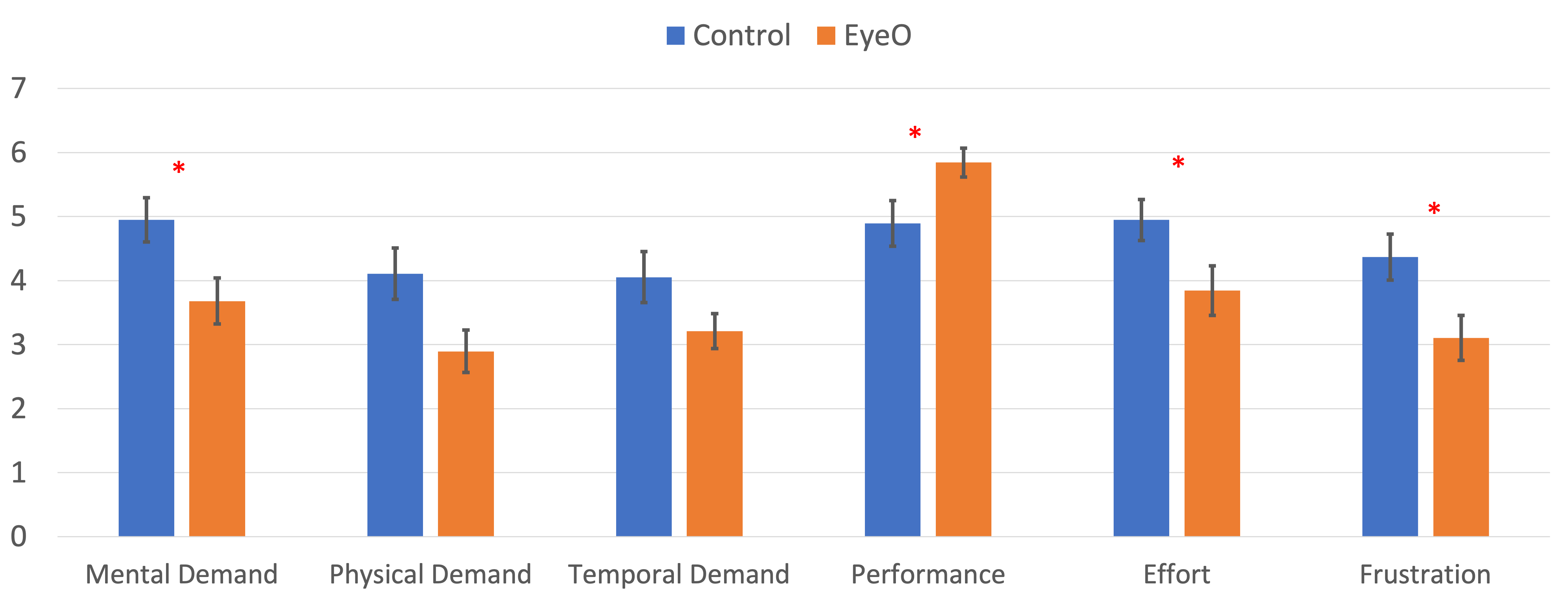}
  \caption{NASA-TLX responses about user workload for the two systems. Significance codes: * < .05, ** < .01, *** <.001}
  \label{fig:nasa-survey}
\end{figure*}

Participants consistently rated \systemName{} more favorably than the control manual calibration in the NASA TLX survey, which assesses a system's mental workload in terms of mental demand, physical demand, temporal demand, performance, effort, and frustration. In Fig.~\ref{fig:nasa-survey}, we observe that along four of the six dimensions, differences between \systemName{} and the static control are statistically significant (p$<0.05$). Participants perceive reduced mental demand (\systemName{}: M=$3.68$, SE=$0.36$; Control: M=$4.95$, SE=$0.35$), improved performance (\systemName{}: M=$5.84$, SE=$0.23$; Control: M=$4.89$, SE=$0.36$), reduced effort (\systemName{}: M=$3.84$, SE=$0.39$; Control: M=$4.95$, SE=$0.32$), and reduced frustration (\systemName{}: M=$3.10$, SE=$0.35$; Control: M=$4.37$, SE=$0.36$).
These results highlight an improved and more seamless user experience with autocalibration via \systemName{} compared to a static calibration approach. Qualitative feedback (Sec.~\ref{sec:pref}) provides further support for these findings along the dimensions of mental comfort, performance, effort, and frustration.

We note that several participants asked us clarifying questions about the questions for physical demand and temporal demand. It is possible that the results for these two dimensions may reflect discrepancies in their interpretations.

\subsubsection{Overall Preferences}
\label{sec:pref}
We asked participants ``Which system do you prefer to use as an end-user of this device?''. Their preferences between the systems were: \systemName{} 14 (73.68\%), control 3 (15.79\%), no preference 2 (10.53\%). Open feedback from participants, summarized below, sheds light on these preferences.

Participants who preferred \systemName{} cited increased comfort and ease. They explained, \emph{``I think [\systemName{}] is more comfortable. With [the control], I had to compensate for each word and found it harder to select the letter''}; \emph{``I prefer [\systemName{}], because it was on average easier and required less mental and physical load.''} Another noted the accuracy of the system, \emph{``As an end-user I prefer [\systemName{}] as it was more accurate for most of the sentences I typed. It was only frustrating for 2/5 sentences, as opposed to [the control] which was frustrating for most of the sentences.''}. Participants also appreciated the real-time updates, one user noting \emph{``[\systemName{}] adapted quickly to where I was looking''}. Others said, \emph{``[\systemName{}] seemed to adjust to my typing and improve as I performed the task whereas  [the control] was constantly bad (and while the consistency helped and I learned to adjust where the system faltered, I was frustrated I needed to apply additional effort to complete the task)''}; \emph{``In [\systemName{}], I would start off having to recognize the offset and type accordingly, but after a few characters it would adapt and then I could actually look at the intended character, so it got progressively easier. In [the control] interface, when it worked (1 case) it was very smooth, but for the other 4 cases I had to identify the offset and consistently use it to type the whole phrase.''} Some users also perceived that their own ability to type improved with \systemName{}, \emph{``When starting with [\systemName{}], I did notice I was better at using the input method (typing with eyes) as I used more sentences, irrespective of how difficult the given sentence was to type as I was more comfortable overall.''}.

Since we purposefully introduced miscalibration to evaluate \systemName{}'s corrections, some users also noted the struggle with \systemName{} before the autocalibration started with the first reading attempt, even though overall they preferred it over the static control. \emph{``I found [the control] to be substantially more difficult to use. I felt significant strain while using it and felt noticeably tired after completing the five [control] tasks.
[\systemName{}] was not easy to use, and required significant focus to complete, but I did not feel as strained while using it.
Overall, I would describe [the control] as difficult to use and very uncomfortable, and [\systemName{}] as difficult to use but only somewhat uncomfortable''}. When asked further to explain their reasoning for preferring [\systemName{}], they noted, \emph{``I much preferred [\systemName{}]. I don't think I could have completed more than five sentences using [the control] without having to take a break. With [\systemName{}] I believe I could have done eight to ten sentences before needing a break. Even with a break, however, I do not think I could complete more than two five-sentence sessions using [the control] without needing to walk away from the computer for an extended (10+ minute) period of time. While answering the survey for [the control], I felt like I had a headache. I was worried that the headache would get worse during the next session of typing, but it did not. [\systemName{}] was not comfortable to use, but it did not cause me the sort of discomfort and mental strain I felt during and immediately after using [the control].''}.

Participants who did not prefer \systemName{} cited some pitfalls of our approach. While most users read the typed text, we noticed one user instead relied on activated keys turning red. If a user does not look at the text box to read the typed text, \systemName{} does not update the calibration. A few users also preferred the reliability of errors in the static calibration system, despite additional cognitive load.
As one participant explained, \emph{``The error or offset between where I gaze and the detected cursor seems constant in [the control]. In [\systemName{}], the error is more random.''} We note that since users were not aware of the autocalibration feature relying on their reading, this contributed to unpredictability for a minority of the users. Though the purpose of the study was to evaluate `implicit' autocalibration, we believe if users were made aware of the underlying autocalibration technique, that could further enhance their experience.

One participant summarized the tradeoff between systems: \emph{``It felt as if [\systemName{}] was adapting to the miscalibration in the eye tracker, whereas [the control] was not. The adaptability of [\systemName{}] has benefits and tradeoffs. It meant that [the control] was more predictable, whereas [\systemName{}] was less predictable. That said, there was perhaps less overall motor coordination effort involved in using [\systemName{}]''}. The same user noted, \emph{``I think that overall I would prefer [\systemName{}] in the long-term.''} 
\section{Semi-structured Interview with Stakeholders}
\label{sec:als-interview}

To gain a better understanding of the potential usefulness of our autocalibration method for people with disabilities who use gaze typing, we conducted a semi-structured interview with stakeholders from the ALS community.

\subsection{Procedure}
We ran a single virtual semi-structured interview session with a group of ALS stakeholders, with IRB approval. The session consisted of (1) a demonstration of \systemName{} and (2) a guided group discussion. The group discussion centered around participants' relationship with the ALS community and gaze typing, barriers to gaze typing technologies for ALS users, resources to assist ALS users during gaze typing, and feedback on the demoed \systemName{} prototype. The discussion was transcribed for further analysis.

Two authors analyzed the transcript via deductive thematic coding \citep{braun2006using}. 
Three primary themes were derived from the semi-structured interview questions and any additional content covered during the interview. Each primary theme was subdivided into multiple secondary themes.
The transcript was categorized sentence-by-sentence into one or more secondary codes, if applicable. During the coding process, three authors met regularly to discuss the coding assignments. All authors discussed the findings of the thematic coding, which we summarize  below.

\subsection{Participants}

We recruited 7 participants from support organizations for ALS. 
Two of the participants were direct caregivers for family members who have/had ALS, three were members of organizations which support people with ALS, 
and one was a speech pathologist who treats ALS clients in the states of Washington, New Jersey, and Maryland in the United States. One of the participants has 20 years of experience in software and hardware development, with a focus on accessibility programming and building eye trackers. Another has 40 years of software development experience and got involved in the ALS association 18 years ago, building software for missing pieces of eye tracking to make them do things they weren't programmed or capable of doing at the time. In addition, we were joined by a participant with cerebral palsy who used eye trackers for gaze typing in her daily life. 

In the ALS community, caregivers are heavily relied on for setting up and using gaze tracking or other assistive devices. In particular, support staff who specialize in assistive devices and care for people with ALS have a wealth of experience from working with many individuals, and we aimed to learn from their broad experience with many people.

\subsection{Resulting Themes}
We discuss our findings from the semi-structured interview categorized by the three identified primary themes and their subsequent secondary themes.

\subsubsection{Obstacles to Eye-Gaze Typing} 
Participants discussed various obstacles ALS users face in using eye tracker for gaze typing, which is an essential form of communication.

\begin{itemize}
    \item \textbf{Medications:} Participants noted that during the progression of ALS, people ingest stronger doses of medications such as muscle relaxants, anti-allergens, and opioids to manage symptoms and discomfort. Each of these three categories of medications creates problems for gaze-based interaction. Muscle relaxants make it harder to focus on a precise location on the screen; anti-allergens cause eye dryness which induces blinking, thereby diminishing eye tracking quality; and opioids cause pupil constrictions and dilations, making it harder to track the eyes. These issues 
    could also create autocalibration problems for users with progressed disease.

    \item \textbf{Eye Function:} Participants noted that glasses often diminish eye tracking quality as they reflect both external light sources and the screen itself. 
    Thus, many people with ALS do not wear glasses during gaze typing. Additionally, towards later stages of the disease, one side of the body often tightens, causing the head to drop and rotate to one side and making eye tracking more difficult. As a result, in later phases of the disease, only one eye will be used for tracking. Most modern eye trackers (including Tobii PCEye, used in our work) can reliably detect gaze by tracking only one eye.

    \item \textbf{Changing Positions:} Participants shared that people with ALS often move out of their chair or have their screens removed for medical care. Every time they return, they must recalibrate their eye tracker.  
    They suggested that autocalibration could benefit such users returning to their device after intermissions without having to recalibrate each time. One participant explained, \emph{``When tracking errors are position specific... an autocalibration algorithm that was the right one, that would be very useful and would operationalize eye trackers properly.''}
    
    \item \textbf{Calibration Process:} 
    Participants shared that ALS users often adapt and find workarounds for calibration drift, excelling at compensating for miscalibrations. One care worker explained, \emph{``Some of their brain muscle memory has already developed like they know how they can be so off and all of a sudden they shoot to the right character and get it. And it's because they've learned how to adapt to that.''}

    \item \textbf{Recalibration Frequency:} Participants explained that ALS users often recalibrate as much as 50 times a day, in an attempt to fix poor tracking. This causes frustration and can discourage use of the device at all. \emph{`` It's annoying to them and they need to calibrate again and again'; `..they'll need to have 20 calibrations in a 2 hour period. Well, there's no solution other than recalibrating. So like this would be amazing, them having the ability to have that autocalibration.''}

\end{itemize}

\subsubsection{Resources for Eye-Gaze Typing}
Participants discussed resources that help alleviate obstacles to gaze typing for people with ALS. 
\begin{itemize}
    \item \textbf{Human debugging:}
    Participants noted that gaze typing can cause fatigue and overwhelm ALS users new to the system. Keyboard layouts are modified over time with additional features or buttons to help them use the device for longer and make the learning curve easier.
    In debugging, feedback from calibration (e.g. patterns in mistakes across the calibration targets) is a tool that care givers use. 
    For example, such feedback can indicate that the device positioning is off, or that lighting from a window is interfering with tracking 
    \emph{``There are so many situations where you need someone to figure out what you see from the calibration''}. 
    \item \textbf{Education and training:} Participants noted that there is a lack of awareness and education 
    about using eye tracking technology effectively with ALS. For example, they noted that education about the impacts of medications could significantly advance eye tracking efficacy.
\end{itemize}

\subsubsection{Prototype Feedback}
Finally, participants shared their perspectives on the demoed prototype's potential, pitfalls and suggested future improvements for ALS users.
\begin{itemize}
    \item \textbf{Use Cases:} Participants found \systemName{} to be promising for ALS users, and were interested in ALS users having the opportunity to use the system. \emph{``We would definitely try it with our clients 100\%.''} In particular, they thought that \systemName's autocalibration could help alleviate the frustration around repeated calibration. \emph{``It would be amazing for them to have the ability for autocalibration.''} 
    They also recognized the potential for \systemName{} to reduce the learning curve for new users and improve communication during the progression of ALS. \emph{``The trackers work really well on some [people] and then they get towards the more progressed stage of the disease, towards the end of life, and then that's when a lot of the times things get complex with medications. So I'm just really happy to see anything that could help get somebody started on eye gaze, because  a lot of people start too late.''}; \emph{`'For someone who's starting off and they have a significant offset, this would be very, very significant.''}
    \item \textbf{Desired Improvements:} Participants noted that for people in later stages of ALS or people who are otherwise unable to dwell on a precise location, having larger keys areas and fonts could help increase the error tolerance of the algorithm and still prove helpful. 
    They suggested that specific keys could further enhance autocalibration -- for example, the spacebar could serve as an indicator that the previous word was typed correctly, or typing the next likely character successfully could be used as a point of calibration. They also suggested that alternate sensory feedback during calibration such as audio could increase transparency and enhance the user experience by letting the user know that the calibration is changing. \emph{``I think even being able to alert the user to what has changed specifically before and after autocalibration would be really important.''} 
    They also noted that logs or reports generated during autocalibration could be useful for caregivers debugging the system and for physicians diagnosing other medical conditions like cataracts or medication effects.  
    
    \item \textbf{Pitfalls:} Participants reiterated that people using medications which cause dry eyes or pupil dilations might not benefit as much from autocalibration because their main challenge with gaze typing is the inability to focus or the eyes not being detected by the eye trackers. They also mentioned that miscalibrations can differ across the screen, but our work only calibrated to a fixation point on the top of the screen.
    \item \textbf{Control to turn autocalibration on/off:} Participants recommended providing the option to turn autocalibration on or off, and suggested that  
    AI could recommend to the user when to turn it on. \emph{``The one thing that would concern me would be breaking things for people that aren't having a problem. You know, someone whose brain muscle memory has already developed [for compensating to miscalibration]. One suggestion would be the system detects things are not working great for them and an AI [tells] them that they're a little bit off. The system could suggest OK, you're not doing quite as well as you were before. Let's do a little bit of autocalibration to see if we can get you back on track?''}
\end{itemize}
\section{Discussion}
\label{sec:discussion}

As this work suggests, autocalibrating eye trackers without introducing additional tasks for the end user represents a step towards building more seamless user experiences. Gaze data offers vast potential as a form of input for computer interaction, effectively making technology more accessible and user-friendly. Coupled with methods for autocalibration, an array of seamless gaze-controlled interactions may become possible in the future.
To help inform such developments, below we discuss how our work informs gaze control more broadly, outline future work, and discuss our work's limitations.

\subsection{Limitations}
While our technique introduces new possibilities for making interactions more seamless, it also has several limitations. As in gaze tracking generally, our technique relies heavily on users having clear, unobstructed vision and the capacity to make smooth eye movements across the visual keyboard, which may not always be the case, particularly for users with specific ocular or neurological conditions. Additionally, eye movements during reading and writing are influenced by numerous cognitive and physiological factors, such as attention, fatigue, and cognitive load. These factors can cause dynamic changes in eye movement behavior, potentially affecting the computation of the miscalibration offset and hence the calibration accuracy. 

We also recognize that our user study did not include people with disabilities or others who regularly rely on gaze typing, though our semi-structured interview engaged more with gaze typing stakeholders. 
While our user study results suggest that the proposed autocalibration approach can significantly improve the typing efficiency and user experience for able-bodied participants, the comfort, ease-of-use, and adaptation to individual user needs for disabled populations are factors that must be evaluated and explored further in future work.

\subsection{Implications for Gaze Control}

The insight that gaze behavior differs when used for input (e.g. typing) and output (e.g. reading) can be leveraged for gaze interactions beyond gaze typing. 
Any interactive application where gaze functions as control at some points but is used for perception at other points can leverage this insight to improve performance and calibration. For example, an AR/VR meeting that enables participants to select from a menu through gaze control, and also provides captions in another part of the screen, autocalibration can safely occur while the user reads the captions to improve interaction with the menu and other parts of the application. 
Applications in VR/AR for gaze-based web browsing, gaming, and computer-control 
can especially benefit in terms of user experience and control latency with precise gaze prediction.

We also note that the insight we leverage about differences in gaze behavior is closely related to the Midas Touch problem. The Midas Touch problem \citep{velichkovsky1997towards} states that gaze used for both perception and control can trigger unintended actions (such as activation of unintended keys while typing). Until now, this problem has served as a barrier to seamless gaze-controlled interactions. We note however, that when the differences in gaze behavior can be well isolated to different parts of an application -- i.e. if the system can accurately predict whether the user is engaging in control or perception -- the system can strategically leverage this difference. In our work, we leveraged this difference to selectively trigger calibration processes. This difference could also be leveraged to selectively trigger control events only when the user is attempting control, and prevent control events when the user is engaging in perception. 
It may also be possible to design interfaces where it is easier to predict if the user is engaging in control or perception. For example, designing visual targets to be used exclusively either for control or perception could boost prediction accuracy of user intent.

\subsection{Future Work}

Future work includes exploring autocalibration for multimodal interfaces and more immersive environments. 
One user study participant shared their excitement for integrating gaze typing technology with other modalities (e.g.  speech or joystick input to smart TVs). They also recognized that the additional cognitive load may necessitate autocalibration and additional support for real-world deployment: \emph{``As someone who is open to alternatives means of typing due to physical restrictions, I like the potential of the technology but also acknowledge the potential strain for persons mentally and physically. It would be cool for persons typing on TVs though''}. Developing and studying how to best autocalibrate and support such multimodal interfaces makes interesting future work.

During the semi-structured interview, participants suggested many potential improvements to autocalibration. For example, their suggestion to improve transparency around when autocalibration occurs introduces rich HCI research opportunities. Future studies can explore what information should be made available to the user (e.g. binary signal to indicate autocalibration, detected errors indicating miscalibration, and/or the direction of correction), and the interface design for conveying this signal (e.g. auditory/ visual feedback). 

Additionally, designing a feature to enable user control for autocalibration introduces interesting research questions for future work -- What amount of detected miscalibration should trigger suggestions for autocalibration? Should these recommendations be personalized? Can the impact of user tolerance to miscalibrations/compensation be learned?
While we tested our autocalibration prototype with a single fixation point on the text box placed at the top of the screen, additional perception-based landmarks could be placed on different parts of the screen to help estimate a fine-grained miscalibration map. Prior approaches using saliency maps, correlation of smooth pursuits and moving targets, next word predictions etc. can further augment \systemName{} for enhanced autocalibration, and in turn more seamless gaze interactions.

Furthermore, as we move towards a future where human-computer interaction becomes increasingly multifaceted, incorporating a higher bandwidth of input signals like speech, gestures, and facial expressions becomes essential. The autocalibration approach, by enhancing the accuracy and efficiency of eye tracking, can contribute towards creating more holistic, powerful, and intuitive interfaces especially for real-world applications like augmented reality (AR) and virtual reality (VR). With increasing reliance on these technologies for various purposes – from gaming to professional training – the need for seamless and efficient interfaces is paramount. Our work can contribute towards this goal by offering more accurate eye-tracking, thereby improving the user's interaction with the AR/VR environment. %
\section{Conclusion}
\label{sec:conclusion}

Eye tracking technology has enabled computer interactions for many users, with the potential to unlock additional future applications and interactions. 
Gaze typing is a common application of eye tracking technology, and is particularly beneficial to those with physical disabilities who cannot use traditional keyboards. 
However, the accuracy and efficiency of gaze tracking (including gaze typing) can be significantly impacted by the calibration of the eye tracking system. Conventional calibration methods are time-consuming and require users to periodically perform manual calibrations, which can be inconvenient and disruptive to the user experience. Users also often compensate for miscalibration by intentionally ofsetting their gaze, confounding the automatic detection and correction of miscalibration. As a result, calibration problems remain a significant barrier to use.

To help address this problem, we provide the insight that gaze is used for both input and output, and differences in performance between these two modes can enable detection and correction of miscalibration. We demonstrate this approach through our \systemName{} prototype, which provides autocalibration for gaze typing to provide a more seamless and accurate user experience. Our approach dynamically compensates for miscalibration using the difference in users' gaze during typing (when the user may compensate for miscalibration) versus reading the text they have typed (when the user does not compensate). Results from our user study suggest that by leveraging the natural gaze behavior of users during reading, our autocalibration system provides a more efficient, less mentally demanding, and overall preferable experience compared to traditional manual calibration. Results from our semi-structured interview with stakeholders deepens understanding of users' difficulties, resources, and the potential utility and future improvements for such autocalibration systems. 
We hope that other researchers and developers find useful insights in our work, and in particular are able to similarly leverage the difference in gaze performance during input and output to improve user experiences.

\section*{Acknowledgments}
The authors would like to thank Pete Ansell, Jay Beavers, Jon Campbell, Susan Dumais, Harish Kulkarni, John Tang, and Shane Williams for helpful discussions and feedback.

\bibliography{main}
\bibliographystyle{abbrvnat}

\end{document}